\documentstyle[11pt,newpasp,twoside,epsf]{article}
\def\edcomment#1{\iffalse\marginpar{\raggedright\sl#1\/}\else\relax\fi}
\reversemarginpar

\begin{document}
\title{Discovery of Active Galactic Nuclei in Mid- and Far-Infrared
       Deep Surveys with ISO}
\author{Yoshiaki Taniguchi}
\affil{Astronomical Institute, Graduate School of Science,
        Tohoku University, Aramaki, Aoba, Sendai 980-8578, Japan}

\begin{abstract}
We present a summary on the discovery of active galactic nuclei
in mid- and far-infrared deep surveys with use of the Infrared
Space Observatory.
\end{abstract}

\section{Introduction}

Active galactic nuclei (AGNs) such as quasars have been
one of important issues of modern astrophysics
since the discovery of a quasar 3C 273 in 1963 (Schmidt 1963).
Their central engine has been thought to be 
mass-accreting, super-massive single black holes around which
gravitational energy is transformed into huge kinetic and radiation
energy with the help of gaseous accretion disks (e.g., Rees 1984).
Since AGNs are intrinsically bright in any frequency regime,
various kinds of AGN surveys have been conducted by using radio through 
optical to X-ray telescope facilities (e.g., Peterson 1997). 

Since current unified models for AGNs introduce a dusty torus
around the central engine (see for reviews, Antonucci 1993;
Urry \& Padovani 1995), any infrared information is absolutely 
necessary to understand AGN phenomena because such dusty tori
radiate anisotropic infrared emission (e.g., Pier \& Krolik 1992;
Clavel et al. 2000; Laurent et al. 2000; 
Murayama, Mouri, \& Taniguchi 2000 and references therein). 
Another important issue related to infrared observations of AGNs
seems to be possible starburst-AGN connections (Weedman 1983; 
Sanders et al. 1988; Heckman et al. 1989;  Mouri \& Taniguchi 1992,
2001; Cid Fernandes et al. 2001; Storchi-Bergmann et al. 2001).
In particular, infrared emission properties are highly useful  
in investigating circumnuclear star formation and its relation
to the nuclear activity in AGNs (e.g., Genzel et al. 1998; Tran
et al. 2001).

Most Seyfert galaxies and a part of quasars were indeed detected
at a wavelength range between 12 $\mu$m and 100 $\mu$m
by the Infrared Astronomical Satellite, {\it IRAS}, launched
in 1983  (see for a review, Soifer, Houck, \& Neugebauer 1987).
Although {\it IRAS} all-sky survey covered over 96\% of the sky,
its completeness flux limit was 0.5 Jy at 12 $\mu$m, 25 $\mu$m,
and 60 $\mu$m, and 1.5 Jy at 100 $\mu$m. Therefore, {\it IRAS}
could probe infrared galaxies and AGNs up to redshift $\sim$ 0.2
-- 0.3 (e.g., Veilleux et al. 1999; Borne et al. 2000)
except for some unusually bright (or gravitationally 
amplified) sources beyond redshift $z \sim 0.5$, e.g., 
IRAS F10214+4724 at $z = 2.28$ (Rowan-Robinson et al. 1991) and
IRAS F15307+3252 at $z = 0.93$ (Cutri et al. 1994);
see also Rowan-Robinson (2000).

More than 10 years after the launch of {\it IRAS}, the Infrared
Space Observatory, {\it ISO}, was launched in 1994 (Kessler et al.
1996). The two infrared array detectors, ISOCAM (Cesarsky et al. 1996)
and ISOPHOT (Lemke et al. 1996), were used to carry out mid-infrared
(MIR) and far-infrared (FIR) deep surveys, respectively.
In this paper, we will give a summary of the MIR and FIR deep surveys 
with {\it ISO}
and then describe what kinds of AGNs are newly found in these surveys
(see for a review, Genzel \& Cesarsky 2000).

\section{MIR Deep Surveys with ISO}

ISOCAM has two independent channels containing each a 32$\times$32
pixel detector; 1) the short wavelength channel (2.5 $\mu$m to 
5.5 $\mu$m) and the long wavelength one (4 $\mu$m to 18 $\mu$m)
(Cesarsky et al. 1996). Since the short wavelength can be accessible
from ground-based telescope facilities, the long wavelength camera
was used to carry out MIR deep surveys; i) at 7  $\mu$m (Taniguchi et al.
1997; Sato et al. 1999, 2001; Rowan-Robinson et al. 1997; Aussel et al.
1999; Oliver et al. 2000; L\'emonon et al. 1998; Altieri et al. 1999;
Fadda et al. 2000), ii) at 12  $\mu$m (Clements et al. 1999), and
iii) 15 $\mu$m (Rowan-Robinson et al. 1997; Aussel et al. 1999;
Oliver et al. 2000; L\'emonon et al. 1998; Altieri et al. 1999;
Fadda et al. 2000). A summary of these surveys is given in Table 1;
note that the 12 $\mu$m survey by Clements et al. (1999) is not included
in this table.

\begin{table}
\caption{A summary of the MIR surveys with ISO}
\begin{tabular}{cccccc}
\tableline
\tableline
Field\tablenotemark{a} & Area\tablenotemark{b} & 
$F_{\rm lim}$(7$\mu$m)\tablenotemark{c} & $N$(7$\mu$m)\tablenotemark{d} &
$F_{\rm lim}$(15$\mu$m)\tablenotemark{e} & $N$(15$\mu$m)\tablenotemark{f} \\
 &  (arcmin$^2$) & ($\mu$Jy) &  &
 ($\mu$Jy) &  \\
\tableline
HDF-N$^1$ & 5/9 & 65 & 7 & 200 & 45 \\
LHNW$^2$  & 9/--- & 32 & 27 & --- & --- \\
SSA 13$^3$ & 16/--- & 6 & 65 & --- & --- \\
CFRS$^4$ & ---/100 & --- & --- & 250 & 78 \\
ELAIS$^5$ & 11.7/22.8\tablenotemark{g} & 1000 & $\sim$ 700 & 2000 & $\sim$ 800 \\
A2390C$^6$ & 5.76/5.76 & 65 & 4 & 65 & 4 \\
A2390$^7$ & 6.76/6.76 & 25 & 31 & 40 & 34 \\
A1689$^8$ & 36/36 & 150 & 41 & 300 & 18 \\
\tableline
\tablenotetext{a}{HDF-N = Hubble Deep Field-North, LHNW = the NW field
in the Lockman Hole, SSA 13 = Hawaii Small Selected Area No. 13,
CFRS = Canada-France Redshift Survey Field, ELAIS = European Large Area
ISO Survey field, A2390C = the central region of Abell 2390, A2390 =
Abell 2390, \& A1689 = Abell 1689. Their references are;
1. Aussel et al. 1999, 2. Taniguchi et al. 1997, 3. 
Sato et al. 1999, 2001, 4. Flores et al. 1999, 5. Oliver et al. 2000, \&
Serjeant et al. 2000, 6. L\'emonon et al. 1998, 7. Altieri et al. 1999,
\& Fadda et al. 2000.}
\tablenotetext{b}{Sky coverage. The first and second numbers means the sky coverage
                  at 7$\mu$m and 15 $\mu$m, respectively.}
\tablenotetext{c}{The 3 $\sigma_{\rm rms}$ flux limit at 7 $\mu$m in the survey.}
\tablenotetext{d}{The number of objects detected at 7 $\mu$m in the survey.}
\tablenotetext{e}{The 3 $\sigma_{\rm rms}$ flux limit at 15 $\mu$m in the survey.}
\tablenotetext{f}{The number of objects detected at 15 $\mu$m in the survey.}
\tablenotetext{g}{Note that the area for ELAIS is given in units of square degree.}
\end{tabular}
\end{table}

\begin{table}
\caption{AGNs found in the MIR surveys with ISO}
\begin{tabular}{cccccc}
& \\
\tableline
\tableline
Field\tablenotemark{a} &
$N_{\rm gal}$\tablenotemark{b} &
$N_{\rm AGN}$\tablenotemark{c} & $f_{\rm AGN}$\tablenotemark{d} &
$N_{\rm Type1}$\tablenotemark{e} & $N_{\rm Type2}$\tablenotemark{f} \\
 &  &  & (\%) &   &  \\
\tableline
HDF-N & 49 & 1 & 2 & 1 & 0  \\
LHNW  & 13 & 1 & 8 & 1 & 0  \\
CFRS  & 78 & 11 & 14 & 4 & 7 \\
A2390 & 32 & 0 & 0 & 0 & 0 \\
A1689 & 45 & 1 & 2 & 1 & 0 \\
\tableline
Total & 217 & 14 & 7 & 7 & 7 \\
\tableline
\tablenotetext{a}{The same as those in Table 1.}
\tablenotetext{b}{The number of galaxies detected in the survey.}
\tablenotetext{c}{The number of AGNs detected in the survey.}
\tablenotetext{d}{The fraction of AGNs = $N_{\rm AGN}$/$N_{\rm gal}$.}
\tablenotetext{e}{The number of Type 1 AGNs detected in the survey.}
\tablenotetext{f}{The number of Type 2 AGNs detected in the survey.}
\end{tabular}
\end{table}

Among the above MIR imaging surveys, follow-up spectroscopy has been done 
for five surveys given in Table 2. Eleven AGNs (4 type 1 and 7 type 2 AGNs)
were found among 78 sources in the CFRS-ISOCAM survey (Flores et al. 1999).
However, only a few AGN were found in the remaining four surveys (Aussel et al. 1999;
Taniguchi et al. 1997; Taniguchi 1999, 2000; Altieri et al. 1999; Fadda et al. 2000). 
In total, only 14 AGNs (7 type 1s and 7 type 2s) were found among 217 sources
in the five surveys. This gives an AGN
fraction, $f_{\rm AGN} = N_{\rm AGN}/N_{\rm gal} \approx$ 7\%.
In Figure 1, we show an optical spectrum of a quasar at $z=1.025$
found by Taniguchi et al. (1997) as an example.

Although the number of AGNs found in the MIR surveys with {\it ISO} is not so large,
it seems interesting to compare the above observational result 
with some model predictions.
Oliver et al. (1997) estimated expected numbers of AGNs for their MIR survey
of HDF (see also Rowan-Robinson et al. 1997; Aussel et al. 1999), adopting 
the following two models; 1) PRR models (Pearson \& Rowan-Robinson 1996),
and 2) AF models (Franceschini et al. 1994);
in this article, we do not give details of both PRR and AF models.

\begin{description}

\item{
a) AGNs expected in the 7 $\mu$m survey: In the case of PRR models, 
the expected number of AGNs in the ISO/HDF field is 1.26 (0.66 type 1 and 0.6
type 2 AGNs) if the 7 $\mu$m 
limiting flux is 38.6 $\mu$Jy and the AGN fraction is $f_{\rm AGN} \approx$
22\%. On the other hand, in the case of AF models, they are 0.34 and 9\%,
respectively. 
}

\item{
b) AGNs expected in the 15 $\mu$m survey: In the case of PRR models, 
the expected number of AGNs in the ISO/HDF field is 1.51 (0.73 type 1 and 0.78
type 2 AGNs) if the 15 $\mu$m
limiting flux is 255 $\mu$Jy and the AGN fraction is $f_{\rm AGN} \approx$
21\%. On the other hand, in the case of AF models, they are 0.25 and 3.6\%,
respectively.
}

\end{description}
It seems difficult to draw any firm conclusions from the above comparisons between
the observations and models because of small-number statistics,
although we do not see significant inconsistency
between them. Future MIR surveys will give us firmer answers. 

Among the {\it ISO} MIR deep surveys, it has been found that there are
a number of MIR (either 7 $\mu$m or 15 $\mu$m) sources 
without optical/NIR counterparts
(Taniguchi et al. 1997; Aussel et al. 1999; Flores et al. 1999).
These results are summarized in Table 3.
It is shown that $\approx$ 18\% of the MIR sources found in the three
MIR surveys have no
optical/NIR counterparts and $\approx$ 10\% of them above 5$\sigma$
detection have no counterparts. 

We cannot exclude a possibility
that some of such sources may be attributed to unexpected noises.
However, if a starburst galaxy with a mass of $10^{11} M_\odot$
at $z \sim 3$ is heavily reddened (e.g., $A_V \sim 10$), this galaxy
could be detected at MIR but not at 2 $\mu$m (Taniguchi 2000).
Recent submillimeter deep surveys have been finding such
dust-enshrouded high-$z$ galaxies (Hughes et al 1998; Barger et al. 1998;
Smail et al. 1999). Therefore, we cannot also exclude a possibility that
deep NIR surveys may miss a certain part of reddened populations,
causing the underestimate of either star formation density or
nonthermal energy density or both in the universe.
It seems important to keep the presence of such MIR sources
without counterparts in mind in future investigations.

\begin{table}
\caption{MIR sources without optical/NIR counterparts}
\begin{tabular}{cccccc}
& \\
\tableline
\tableline
Field\tablenotemark{a} &
$S/N$ & $N_{\rm gal}$\tablenotemark{b} &
$N_{\rm no}$\tablenotemark{c} & $f_{\rm no}$\tablenotemark{d} \\
 & & & & (\%)   \\
\tableline
HDF-N & $>5\sigma$ & 49 &  7 & 14 \\
      & $>4\sigma$ & 51 & 16 & 31 \\
\tableline
LHNW  & $>5\sigma$ & 13 &  2 & 15 \\
\tableline
CFRS  & $>4\sigma$ & 40 &  4 & 10 \\
      & $>3\sigma$ & 34 &  4 & 15 \\
\tableline
Total &  &  & 187 & 34 & 18 \\
\tableline
\tablenotetext{a}{The same as those in Table 1.}
\tablenotetext{b}{The number of galaxies detected in the survey.}
\tablenotetext{c}{The number of galaxies without optical/NIR counterparts.}
\tablenotetext{d}{The fraction of no-counterparts = $N_{\rm no}$/$N_{\rm gal}$.}
\end{tabular}
\end{table}

\begin{figure*}
\plotone{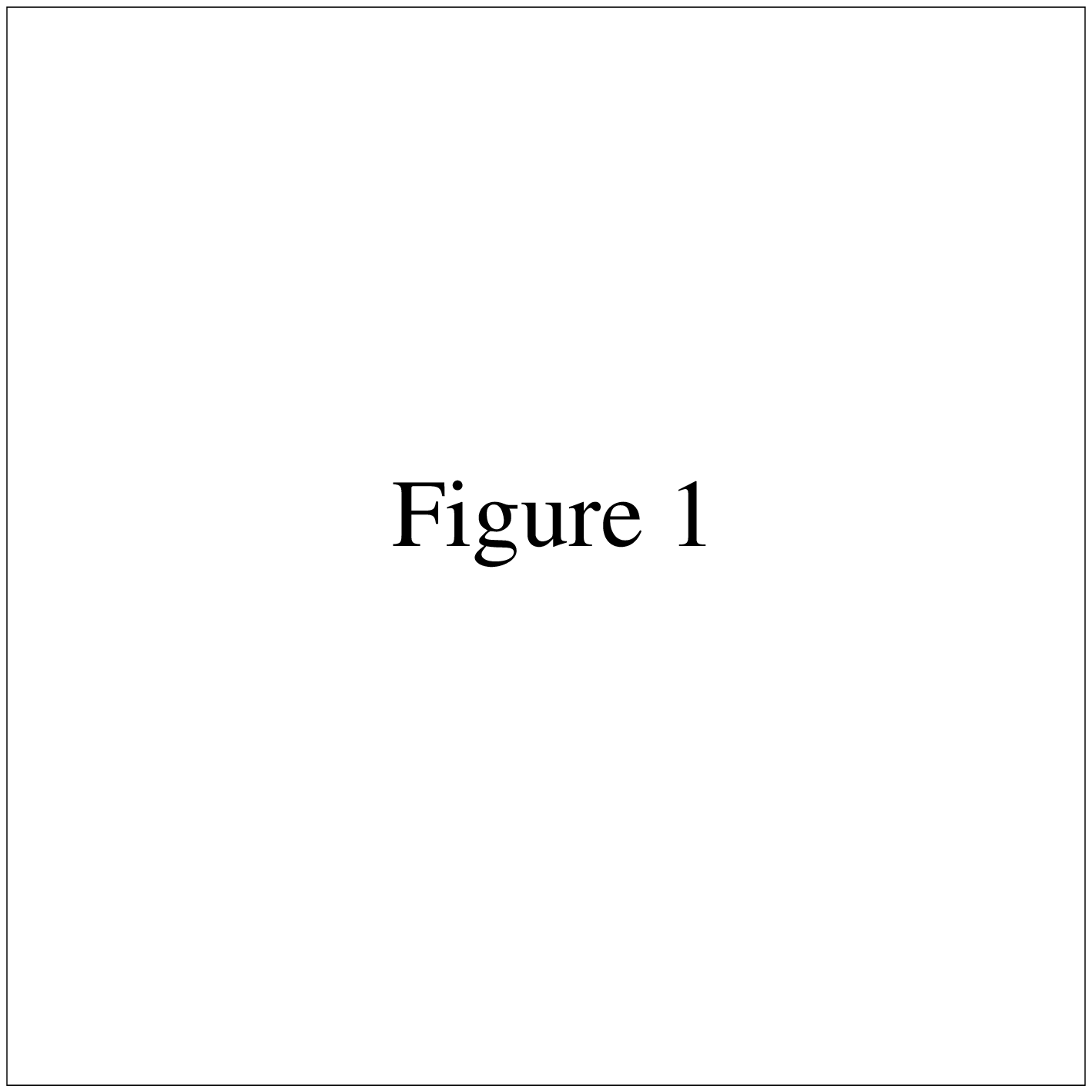}
\caption[]{
Optical spectra of four 7 $\mu$m sources found in the MIR deep survey
by Taniguchi et al. (1997). The spectroscopic observations were made with use
of LRIS on the W. M. Keck 1 telescope (Cowie et al. in preparation).
One quasar at $z=1.025$ was found in their survey (see also Taniguchi 1999).
The middle panel shows the 7 $\mu$m image of the LHNW field (halftone;
darker is brighter). The contours show NIR ($HK$) image taken with the
University of Hawaii 2.2 m telescope. Although the seeing size in the 
NIR image is good (FWHM$\simeq$0.8 arcsec), the image is blurred in 
order to make the comparison between the 7 $\mu$m and NIR images easier.
\label{fig1}
}
\end{figure*}

\newpage

\section{FIR Deep Surveys with ISO}

ISOPHOT is an imaging photopolarimeter covering a wavelength range 
between 2.5 $\mu$m and 240 $\mu$m (Lemke et al. 1996).
Among the several detectors, two detectors, C100 and C200, were
used to carry out FIR deep surveys; i) FIRINDIV-DEEP\&DEEP\footnote{
Contrasting to the FIRBACK-ELAIS program (acronym for Far Infrared
background), in this article,
we call the Japanese-IfA/UH
program FIRINDIV-DEEP\&DEEP where INDIV means individual sources found in
the FIR survey and DEEP\&DEEP means a {\it deep} survey for {\it 
dust-enshrouded extragalactic populations (deep)}. Please make sure that
DEEP\&DEEP is different from the famous DEEP (Deep Extragalactic
Evolutionary Probe) survey program promoted by David Koo at UCO/Lick
Observatory, University of California (http://www/ucolick.org/~deep/home.html).}
(Kawara et al. 1998; Matsuhara et al. 2000), ii) FIRBACK-ELAIS; the 90 $\mu$m
survey (Efstathiou et al. 2000; Serjeant et al. 2001), and the 170 $\mu$m survey
(Puget et al. 1999; Lagache \& Dole 2001; Dole et al. 2001). 
A summary of these survey is given in Table 4; note that we do not 
include another ISOPHOT survey at 60$\mu$m and 90 $\mu$m by 
Linden-Vornle et al. (2000) because no follow-up observation is available.

Although optical followup spectroscopy has not yet been completed for 
the two surveys,  some preliminary results were reported recently.
Serjeant et al. (2001) made optical spectroscopy for 20 sources
detected in the FIRBACK-ELAIS survey with $f$(90$\mu$m) $>$ 100 mJy, 
which are also detected 
either at 15 $\mu$m or at 1.4 GHz. Among them, they found two Seyfert
galaxies at $z=0.149$ and $z=0.225$; the remaining 18 sources are 
either starbursts (16 sources) or early-type galaxies (2 sources).
This gives an AGN fraction, $f_{\rm AGN} = 2/20$ = 10\%. 
On one hand, Kakazu et al. (2001a, 2001b; see also Murayama et al. 2001)
made optical identification of 35 sources found in the FIRINDIV-DEEP\&DEEP
survey with $f$(170$\mu$m) $>$ 100 mJy, using the W. M. Keck, Subaru, UH88, and VLA
facilities. They identified two AGNs; a Seyfert 1.5 galaxy at $z=0.206$
and a quasar at $z=1.60$. The remaining 33 sources are
classified as starbursts (21 sources), LINERs (10 sources), and 
early-type galaxies (2 sources).
Since infrared-selected LINERs may be shock-heated galaxies
(Taniguchi et al. 1999; Lutz, Veilleux, \& Genzel 1999; cf. Imanishi,
Dudley, \& Maloney 2001), 
they are not genuine AGNs but starburst-related (i.e., superwind)
galaxies. Therefore, their survey gives an AGN fraction, $f_{\rm AGN} = 
2/36 \approx$ 6\%.

The above AGN fractions are considered to be tentative values
because the number of observed galaxies are still small.
We hope that future followup observations will give us firmer statistics.
It seems worthwhile noting that Franceschini et al. (1989) estimated
an AGN fraction in such FIR deep surveys; i.e., $f_{\rm AGN} \sim$ 10\%,
being similar to the observed values.

Finally, we mention that 10 ultraluminous infrared galaxies (ULIGs) and 1 
hyperluminous infrared galaxy (HyLIG) were identified as counterparts of
170 $\mu$m sources found in Kawara et al.'s (1998) survey
(Kakazu et al. 2001a, 2001b; Murayama et al. 2001),
Although the HyLIG\footnote{
See for discovery of a HyLIG in the ELAIS survey, Morel et al. (2001).}
is a quasar at $z=1.6$, the ULIGs are located
at $z\simeq$ 0.3 -- 0.8. Such intermediate-$z$ ULIGs have not yet been
found by {\it IRAS}. It is also remarkable that the observed surface density of
ULIGs is $\simeq$ 10 degree$^{-2}$. Note that only $\sim$ 120 ULIGs
with $z < 0.3$ were found by {\it IRAS} (Veilleux et al. 1999; Borne et al. 2000;
see for a review Sanders \& Mirabel 1996), giving a surface density
of $\sim$ 0.003 degree$^{-2}$. Therefore, it is suggested that 
the surface density of intermediate-$z$ ULIGs is much higher by a factor
of 3000 than that of low-$z$ ULIGs. 
Since it has been often argued that ULIGs may be precursors of quasars
(Sanders et al. 1988; Sanders \& Mirabel 1996; Taniguchi, Ikeuchi, \&
Shioya 1999), it is interesting to investigate the nature of 
 intermediate-$z$ ULIGs found in the {\it ISO} FIR deep surveys.
Future wide-field FIR deep surveys and their follow-up observations
will be also very important to find new populations of 
intermediate-$z$ and high-$z$ ULIGs and HyLIGs.

\begin{table}
\caption{A summary of the FIR surveys with ISO}
\begin{tabular}{cccccc}
& \\
\tableline
\tableline
Program &
Field\tablenotemark{a} &
$F_{\rm lim}$(90$\mu$m)\tablenotemark{b} & $N$(90$\mu$m)\tablenotemark{c} &
$F_{\rm lim}$(170$\mu$m)\tablenotemark{d} & $N$(170$\mu$m)\tablenotemark{e} \\
  &  & (mJy) &  & (mJy)  &  \\
\tableline
 FIRINDIV-  & LH          &  45 & 36 & 45 & 45 \\
 DEEP\&DEEP & 1.1 deg$^2$/ &  &  &  &   \\
            & 1.1 deg$^2$ &  &  &  &   \\
\tableline
 FIRBACK-   & Several      & 100(?) & 120(?) & 135 & 106 \\
 ELAIS      & 11.6 deg$^2$/ &  &  &  &  \\
            & 3.89 deg$^2$ &  &  &  &  \\
\tableline
\tablenotetext{a}{LH = Lockman H{\sc i} Hole. As for the FIRBACK survey
                  fields, see http://wwwfirback.ias.u-psud.fr.
                  The first sky coverage is for
                  the 90 $\mu$m survey and the second one is for the 
                  170 $\mu$m one.}
\tablenotetext{b}{3$\sigma_{\rm rms}$.}
\tablenotetext{c}{The number of galaxies detected in the survey.
                  The galaxies detected in FIRINDIV have $F$(90$\mu$m)$>$ 150 mJy.}
\tablenotetext{d}{3$\sigma_{\rm rms}$.}
\tablenotetext{e}{The number of galaxies detected in the survey.
                  The numbers given in this column are that of 
                  galaxies with $F$(170$\mu$m)$>$ 150 mJy for FIRINDIV
                  and that of galaxies with $F$(170$\mu$m)$>$ 180 mJy for FIRBACK.}
\end{tabular}
\end{table}

\section{Discussion and Summary}

As summarized in this article, {\it ISO} enabled us to perform
a number of MIR and FIR deep surveys (see also Genzel \& Cesarsky 2000). 
We give a summary of main points of this article below.

\begin{description}

\item{
1) As for the MIR surveys, the {\it ISO} deep surveys at 7 $\mu$m
and 15 $\mu$m are sensitive enough to detect sources down to
10 $\mu$Jy (Taniguchi et al. 1997; Aussel et al. 1999; Sato et al. 2001b).
Approximately, 10\% of the detected sources appear AGNs from
low-$z$ Seyfert galaxies to a high-$z$ quasar at $z$ 1 -- 2
(Taniguchi 1999; Aussel 1999; Flores et al. 1999).
Since sky areas observed by the surveys are so small,
it seems hard to make statistical arguments on AGN populations.
However, number counts of the MIR sources show significantly stronger 
evolution than what expected from no evolution models
(e.g., Elbaz et al. 1999). This may be attributed mainly to
intense star formation in galaxies at intermediate- and/or high-$z$ 
galaxies (e.g., Takeuchi et al. 2000; Chary \& Elbaz 2001). 
Since dusty starburst galaxies sometimes harbor hidden AGNs 
(e.g., Sanders et al. 1988; Ivison et al. 2000; Willott et al. 2001), 
future follow-up
observations will be important to understand what faint MIR
sources are.
}

\item{
2) The {\it ISO} FIR deep surveys have also shown that 
approximately, 10\% of the detected sources appear AGNs
(Serjeant et al. 2001; Kakazu et al. 2001a, 2001b; Murayama et al. 2001).
Although the total sky area surveyed by the FIR surveys exceeds 10 deg$^2$,
optical follow-up observations have not yet been fully done.
Therefore, future follow-up observations will be important to
understand the nature of faint FIR sources. 
The most remarkable finding of the  {\it ISO} FIR deep surveys
is the discovery of numerous ULIGs at intermediate redshift
between $z \simeq 0.3$ and  $z \simeq 0.8$ (Kakazu et al. 2001a)
because such populations have not yet known from the {\it IRAS}
survey. It seems quite likely that these populations contribute
to the observed excess number count at FIR (e.g., Elbaz et al. 1999;
Takeuchi et al. 2000).
}

\item{
3) The above exciting results urge us to conduct new MIR and FIR surveys.
Indeed, new space infrared telescope facilities will be launched soon;
SIRTF\footnote{http://sirtf.caltech.edu} and IRIS (ASTRO-F\footnote{
http://www.ir.isas.ac.jp/ASTRO-F/index-j.html}).
We hope that these facilities will give us 
much more information on AGN populations in the infrared universe.
}
 
\end{description}
\vspace{5mm}

\acknowledgments{
The author would like to thank the organizers of this nice IAU
colloquium in the beautiful country, Armenia, E. Khachikian, Areg Mickaelian,
Dave Sanders, and Richard Green. 
He would also like to thank his nice colleagues who have been working
together on the MIR/FIR deep surveys with {\it ISO} and their follow-up observations, 
Len Cowie, Dave Sanders, Bob Joseph, Haruyuki Okuda, Kimiaki Kawara,
Yasunori Sato, Toshio Matsumoto, Hideo Matsuhara, Yoshiaki Sofue, 
Ken-ichi Wakamatsu, Youichi Ohyama,
Sylvain Veilleux, Min Yun, Takashi Murayama, Yuko Kakazu,
and Tohru Nagao.}

\end{document}